\documentstyle[prl,epsfig,aps]{revtex}
\newcommand{\be}{\begin{equation}}
\newcommand{\ee}{\end{equation}}
\newcommand{\bq}{\begin{eqnarray}}
\newcommand{\eq}{\end{eqnarray}}
\newcommand{\n}{\noindent}
\newcommand{\nn}{\nonumber\\}

\newcommand{\bc}{\begin{center}}
\newcommand{\ec}{\end{center}}

\newcommand {\al} {\alpha}
\newcommand {\bb}{\beta}

\newcommand{\dd}{\delta}

\newcommand \La {\Lambda}
\newcommand {\OO} {\Omega}
\def \ll {\lambda}

\def\ov{\overline}

 \def\(({\left(}
 \def\)){\right)}
\def\[[{\left[}
\def\]]{\right]}

\def \la{\langle}
\def\ra{\rangle}

\def\Jp{J_{i_1 \ldots i_p}^{l_1 \ldots l_p}}

\begin{document}


\title{{\bf Finite-dimensional corrections to mean field
 in a short-range $p$-spin glassy model}}
\author{Matteo Campellone$^{*}$
, Giorgio Parisi$^{**}$ and Paola Ranieri$^{**}$}

\date{\today}

\address{ (*)
Universitat de Barcelona, Departament de F\'{\i}sica Fonamental, 
Diagonal 647, Barcelona, Spain \\
 (**) Universit\`a di Roma ``La Sapienza''\\
 Piazzale A. Moro 2, 00185 Rome (Italy)\\
 e-mail:{\it  campellone@roma1.infn.it, 
giorgio.parisi@roma1.infn.it, paola.ranieri@roma1.infn.it }}

\date{\today}
\maketitle
\vspace{2truecm}

\begin{abstract}
In this work we discuss a short range version of the $p$-spin model.
The model is provided with a parameter that allows to control the crossover
with the mean field behaviour.
We detect a discrepancy between the perturbative approach and numerical 
simulation. We attribute it to non-perturbative effects due to the 
finite probability that each particular realization of the disorder allows
 for the formation of regions where the system is less frustrated and locally
 freezes at a higher temperature.
\end{abstract}

\vspace{1.cm}
\hspace{.1in} PACS Numbers 05-7510N


\section{Introduction}
The mean field approximation is certainly a very useful starting point
 to study systems with long range interactions. Many physical features may 
differ from mean field behaviour if a model with short range interactions in 
finite dimension is considered.
 Generally speaking, the analysis of the Gaussian fluctuations
is the first step to understand if the mean field 
approximation is well suited to describe the physics of the model. 
The perturbative fluctuations around the minima of the effective
 action may change the critical behaviour of 
the model or even destroy the phase transition. 

In spin glasses and other similar glassy models 
the situation is rather complicated \cite{mpv}.
Generally speaking, one can distinguish two main classes of models:
a first class, which comprehends models that in mean field undergo a
 continuous phase transition with a full replica 
symmetry breaking, and a second class, comprehending 
models which are well described, in mean field, by a 1RSB 
low-temperature solution. 
For this latter class, if the replica symmetry is broken 
with a parameter $q_1$ that is of order $O(1)$,
 the transition is discontinuous.
Models of the first class describe well the physics of real spin glasses
which undergo a second order phase transition with a divergent non-linear 
susceptibility and long range correlations.
For these models
the analysis of the infrared divergences of the propagators 
is rather subtle 
because not all of the modes in replica space are simply related to 
fluctuations of a physical quantity \cite{ddko,fepa}.

In this work we are studying models of the second class where the 
transition does not imply any zero mass mode.
The transition is discontinuous in the sense that, for $T < T_c$,
 another solution, different 
from the high temperature one, becomes dominant 
in the partition function for large $N$.
In this kind of transition there are no precursor effects like 
quantities that diverge as $T$ approaches $T_c$ from above.  
Physically speaking, the discontinuity of the transition reflects
the fact that the number of the metastable states 
becomes non-extensive for low enough temperatures and so a finite probability 
that two replicas belong to the same state arises.
The probability $P(q)$ that two replicas $a$ and $b$ 
have an overlap $q_{ab}=q$ develops
a delta function at $q_1$ when $T < T_c$. The weight of this delta function
 goes to zero as $T$ goes to $T_c$ and it 
increases continuously as the temperature decreases below $T_c$. 
This implies that thermodynamic quantities such as the entropy
 are continuous at $T_c$ and no latent heat is involved in the transition
 which therefore appears to be half way between first and second order. 
One may wonder if the strange features of this transition survive when
 models of the second class are considered in finite dimensions. 

In this work we shall introduce a short range version of the $p$-spin
glass model. The model, defined in $d$ spatial dimensions,
 contains an appropriate parameter $M$ ensuring 
the mean field solution to be exact for $M \to \infty$. 

The analysis of the perturbative fluctuations (which are of order $O(1/M)$)
 around the MF 
solution and of the Gaussian
propagators provides some insight on the nature of the transition in finite
dimension.
A second order phase transition would imply the existence of zero mass 
modes at the critical temperature which reflect the divergence of 
the correlation length.
We will find no divergence of the zero momentum 
propagators and therefore this transition
 seems to invoke no diverging correlations. Moreover, 
 calculating the $O(1/M)$ corrections of the relevant thermodynamic 
quantities around the saddle point, we observe that a perturbative
approach shows no qualitative change of the phase transition with 
respect to mean field: the transition still appears discontinuous 
and still involves no latent heat.

Numerical simulations on the model \cite{cacopa,frpa,papiri} 
nevertheless indicate the existence,
in $d=3$ and in $d=4$, of a divergent susceptibility and of all the 
phenomenology typical of a continuous second order transition at a 
temperature greater or 
equal than the mean field static critical temperature.

This apparent conflict between the two results 
may possibly be explained 
considering non-perturbative effects: the existence of 
regions in space where the system is locally less frustrated than on 
average and in which the spins are long range-correlated even for 
some $T>T_{c}$. 

The paper is organized as follows. In the next section we shall 
briefly review the replica solution of the long-range $p$-spin glass model 
in absence of external magnetic field. We will refer
 to the original references 
for a more detailed discussion of the model and of the replica method.
 In section \ref{srsec} we will introduce our short-range version of 
 the model. In section \ref{gausec} we will calculate the Gaussian 
 fluctuations around the large-$M$ solution of the model in the hot 
 and in the cold phase. In section \ref{prosec} we will calculate the 
 propagators which show no zero-mass modes. We will also discuss the 
propagators on the dynamical line {\em i.e.} obtained by imposing the
 condition of marginality. Finally, we shall draw a possible 
interpretation of the physical behaviour of the model.

\section{The long range $p$-spin model}

The long range $p$-spin glass model is defined by
 the following Hamiltonian

\be
{\cal H}_{p}(\{s\}) =
 -\sum_{(1\leq i_{1}<i_{2}<\cdots<i_{p}\leq N)}\!\!\!\!\!\!\!\!\!\!\!\!\!
J_{i_{1},i_{2},\cdots,i_{p}} s_{i_1}\cdots s_{i_p}  -  h\sum_{i}s_{i},
\label{Hpspin}
\ee

\noindent
where $h$ is an external magnetic field and the interactions 
$J_{i_{1},i_{2},\cdots,i_{p}}$ are random variables 
distributed with the following Gaussian distribution

\be
P(J_{i_{1},i_{2},\cdots,i_{p}}) = \left[\frac{N^{p-1}}{\pi J^2
p!}\right]^{-\frac{1}{2}} \exp{\left[-\frac{(J_{i_{1},i_{2},\cdots,i_{p}})^2
N^{p-1}}{J^2 p!}\right]}.
\label{distrprobp}
\ee
\noindent

The scaling of the variance with $N^{p-1}$ ensures the free energy to be
extensive. For $p=2$ the interactions
 are the more familiar $J_{ij}$ interactions 
and the model is the well known SK model \cite{shki}.
For $p>2$ the model is generally addressed to as the `$p$-spin model'.
The interactions between two spins $s_k$ and $s_l$
 depend on all the $J_{i_{1},i_{2},\cdots,i_{p}}$ having $k$ and $l$
 as one of the arguments as well as all the remaining spin variables of the 
system. 

The behaviour of the model for $p=2$ and for $p>2$ is quite different at low 
temperatures. In this work we shall focus on the physics
of the $p$-spin for $p>2$. 
Note that (\ref{Hpspin}) describes a model which is intrinsically 
mean field for large $N$ since all the $N$ 
spin variables interact reciprocally and there is no geometry in space.

To solve the model, one can introduce $n$ replicas of the system and 
calculate the replicated partition function, \cite{crisom}

\be 
\overline{Z^{n}} = \int \prod \dd J_{i_{1},\cdots,i_{p}}P(J_{i_{1},\cdots,i_{p}})
\sum_{\{s_{i}^{a}\}}\left[\exp{\beta\sum_{a=1}^{n}\left[\sum_{i_{1}<
\cdots < i_{p}} J_{i_{1},\cdots,i_{p}}
s_{i_1}^{a}\cdots s_{i_p}^{a} + h\sum_{i}s_{i}^{a}\right]}\right]. 
\ee

From now on we shall always set $J=1$. 
Integrating over the disorder, and introducing $n(n-1)$ auxiliary fields
$Q_{ab}$ and $\lambda_{ab}$
 (the $Q_{ab}$ and $\lambda_{ab}$ are symmetric $n \times n$ matrices 
with zero diagonal)
 one obtains the following expression
\be
\overline{Z^{n}} = e^{nN\beta^2/4}\int_{-\infty}^{\infty}\prod_{a<b}
dQ_{ab}\int_{-i\infty}^{i\infty}\prod_{a<b}\frac{d\lambda_{ab}}{2\pi}
\exp{\[[-NG(Q_{ab},\lambda_{ab})\]]},
\label{znpspin}
\ee
with
\begin{equation}
G(Q_{ab},\lambda_{ab}) = -\frac{\beta^2}{4}\sum_{a \neq b} Q^{p}_{ab} +
\frac{1}{2}\sum_{a \neq b} \lambda_{ab} Q_{ab} - \ln{Z[\lambda]},
\label{GGG}
\end{equation} 
where
\begin{equation}
Z[\lambda] = Tr_{\{s_{a}\}}\exp\left[\frac{1}{2}\sum_{a \neq b} \lambda_{ab}
s_{a}s_{b} + \beta h \sum_{a}s_{a} \right].
\label{zetaq}
\end{equation}

In equation (\ref{zetaq}) the spin variables have
 lost their dependency from the index $i_r$.
The fields $\lambda_{ab}$ have been introduced as Lagrange multipliers to
impose the conditions 
\be
 Q_{ab} =
\frac{1}{N}\sum_{i}^{N}{s_{i}^{a}s_{i}^{b}}.
\label{qab}
\ee

The saddle point of the functional integral (\ref{znpspin})
 gives the mean field
equations for the fields. Condition (\ref{qab}) gives physical 
meaning to the solution $Q_{{ab}}$ that extremizes 
$G(Q_{ab},\lambda_{ab})$.
The saddle point equations for the $n(n-1)$ fields are

\bq
\lambda_{ab}&=&\frac{\beta^{2} p Q_{ab}^{p-1}}{2},  \nonumber \\
Q_{ab} &=& \overline{<S_{a}S_{b}>} = <S_{a}S_{b}>_{H(Q,S)},
\label{sp1}
\eq

\n
where $<\cdot>_{H(Q,S)}$ stands for the average taken with the measure
 $\exp [-\beta H(Q,S)]$.

If we indicate with $Q^{sp}_{ab}$ and $\lambda^{sp}_{ab}$ the solutions of
equations (\ref{sp1}) the saddle point 
free energy of the system is given by the following formula
\be
F(\beta) = -N\frac{\beta}{4} + \lim_{n\to 0}\frac{N}{\beta n}
G(Q_{sp},\lambda_{sp}).
\label{frs}
\ee

\subsection{Replica Symmetric Solution} 

In the high temperature phase one chooses the RS ansatz for the saddle point 
matrix, so one has

\begin{eqnarray}
Q_{ab} &=& q, \hspace{.3in} \mbox{for $a\neq b$},  \nonumber \\
\lambda_{ab} &=& \lambda \hspace{.3in} \mbox{for $a\neq b$}, \nonumber \\
Q_{aa} &=&\lambda_{aa} = 0.
\end{eqnarray}
The saddle point equations deriving from this ansatz are
\begin{eqnarray}
\lambda&=&\frac{\beta^{2} p q^{p-1}}{2},  \nonumber \\
q &=& \int Dz \tanh^{2}(z\sqrt{\lambda} + \beta h),
\label{sprspspin}
\end{eqnarray}
\noindent
where we used the following notation
\be
\int Dz = \int_{-\infty}^{\infty} \frac{e^{-z^2/2}}{\sqrt{2\pi}}dz.
\ee

In the high temperature phase one has that $q=0$ if $h=0$. 
We shall always consider the $h=0$ case for simplicity
of notations and because 
we shall not be concerned with the effects of a magnetic field.

In this case one finds
\be
F_{RS} = -N\left[\frac{\beta}{4} + \frac{\ln{2}}{\beta}\right].
\label{frs2}
\ee
\noindent
This value of the free energy coincides with the `annealed' result {\em i.e.}
what would be obtained by calculating $\ln{\ov{Z}}$.
From (\ref{frs2}) one can derive the following 
expression for the entropy 
\be
S_{RS} = N \left[ \ln{2} - \frac{\beta^2}{4} \right].
\ee 
\label{srs11}

The entropy in (\ref{srs11}) becomes negative for 
$T < 1/2\sqrt{\ln{2}} \stackrel{\rm def}{=} T_{c}^{*}$,
 so at a greater or equal 
temperature $T_c(p)$ the replica symmetric solution will stop to be correct.
There will be another solution dominating $\overline{Z^{n}}$
 in the zero-$n$ limit, and this will involve the breaking of the original 
symmetry
 among the replicas. The RS solution though, will not become unstable (if 
$p >2$) at $T_{c}(p)$ so the new solution shall be distant from the old one 
and the transition will therefore be discontinuous in the order parameter.

\subsection{1RSB Solution}

Here we show the 1RSB solution for the $p$-spin model for general $p$.
For $p>2$ this solution is correct below 
$T_c(p)$. For finite $p$ there is a second critical temperature 
$T^{*}(p)$ 
at which the system undergoes to a continuous full replica symmetry breaking.
Here we will not study this second transition. So one has, for $p>2$,
$$0<T^{*}(p)<T_c(p)<1.$$

In the large-$p$ limit one has that $T^{*}(\infty) \to 0 $ and
 $T_c(\infty) \to 1/(2\sqrt{\ln{2}}) $. 
The 1RSB solution gives the following expression for $G_{sp}$

\begin{eqnarray}
\lim_{n\rightarrow 0}\frac{G_{sp}}{n} &=&  \frac{\beta^{2}}{4}\left(
mq_{0}^{p} + (1-m)q_{1}^{p}\right)
-\frac{1}{2}\left(mq_{0}\lambda_{0} + (1-m)q_{1}\lambda_{1}\right)
 +\frac{\lambda_{1}}{2} 
 \nonumber  \\
\nonumber \\
 &&-\frac{1}{m} \int D(z) \ln \int D(y)
(2\cosh[\sqrt{\lambda_{0}}z + \sqrt{\lambda_{1}-\lambda_{0}}y ])^m ,
\label{UU}
\end{eqnarray}

\noindent
and the saddle point equations 
 
\bq
&\lambda_{i}&= \frac{\beta^{2} p q_{i}^{p-1}}{2} \nonumber  \\
&q_1& = \int Dz \int Dy \tanh^{2}(z\sqrt{\lambda_0} + y \sqrt{\ll_1 - \ll_0} )
\cosh^m (z\sqrt{\lambda_0} + y \sqrt{\ll_1 - \ll_0} ) \nonumber \\
&m& \frac{\beta^2}{4}(q_1^p - q_0^p)(1-p) = \frac{1}{m}\int Dz\ln \int Dy
\cosh^m (z\sqrt{\lambda_0} + y \sqrt{\ll_1 - \ll_0} ) - \nonumber \\
& \int Dz & \frac{\int Dy \cosh^m (z\sqrt{\lambda_0}
 + y \sqrt{\ll_1 - \ll_0} )
\ln \cosh (z\sqrt{\lambda_0} + y \sqrt{\ll_1 - \ll_0} )}
{ \int Dy \cosh^m (z\sqrt{\lambda_0} + y \sqrt{\ll_1 - \ll_0} )}.
\label{sp1rsbpspin}
\eq

In absence of external magnetic field $q_0=0$. 
$m$ is an additional parameter which 
characterizes the form of the solution. 
The saddle point solution $m_{sp}(T)$ has a physical meaning in the replica
 method. Two different replicas have an overlap $q_0$ with probability 
$m_{sp}(T)$ and overlap $q_1$ with probability $1-m_{sp}(T)$.
 For $0< m_{sp}(T) < 1$ the $1RSB$ solution is the correct one and the 
critical temperature is given by the equation $$m_{sp}(T_{c})=1.$$ 

In the SK model ($p=2$) the RS solution 
becomes unstable at a higher temperature ($T_c^{SK}=1$ 
with our normalizations) and there is a transition into a full $RSB$ phase
with a continuous solution $q(x)$ which holds for all temperatures 
below $T_c^{SK}=1$. In this case the 1RSB is anyhow a better approximation 
than the RS one.

\begin{figure}[htpb]
\centerline{\epsfig{figure=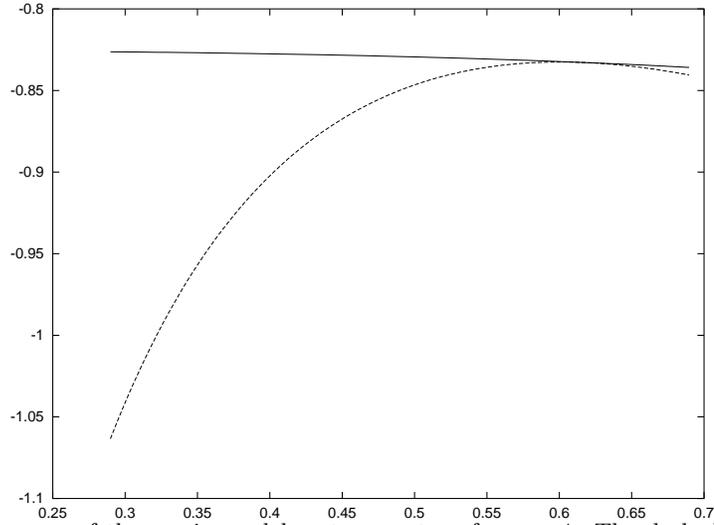,angle=270
,width=10cm}}
\vspace{0cm}
\caption[]{Free energy of the $p$-spin model vs temperature for $p=4$.
The dashed line is the RS solution which is wrong at low 
temperatures.}
\label{f}
\end{figure}

\begin{figure}[htpb]
\centerline{\epsfig{figure=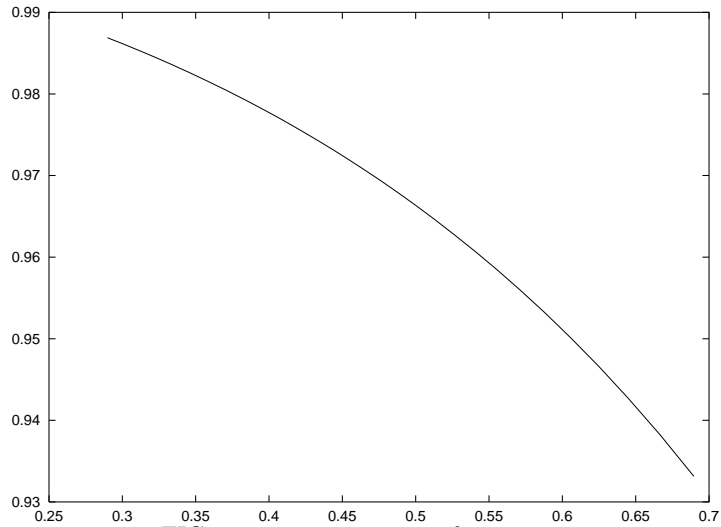,angle=270
,width=10cm}}
\vspace{0cm}
\caption[]{$q_1$ vs temperature for $p=4$.}
\label{q1}
\end{figure}

\begin{figure}[htpb]
\centerline{\epsfig{figure=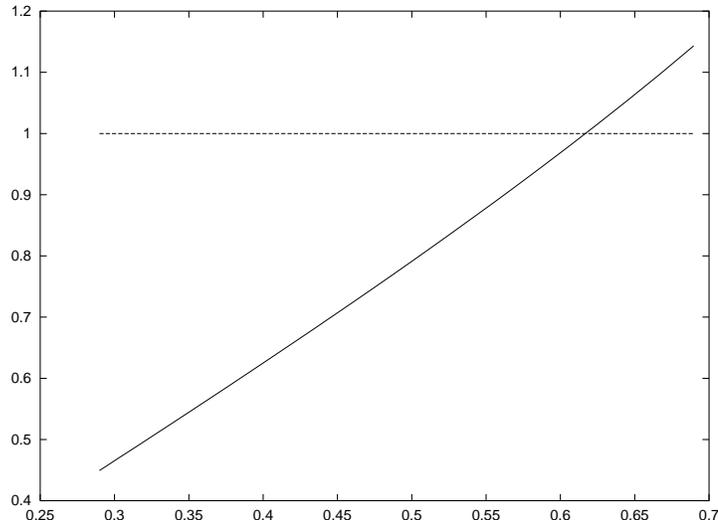,angle=270
,width=10cm}}
\vspace{0cm}
\caption[]{$m$ vs temperature for $p=4$. At the critical temperature 
the curve crosses the line $m=1$.}
\label{m}
\end{figure}

In figure (\ref{f}) we plotted the free energy of the model for $p=4$ for
 low temperatures. 
The high and low temperature curves are tangent at the 
critical temperature.
In figures (\ref{q1},\ref{m}) we plotted the values of $q_1$ and $m$
obtained by the $1RSB$ solution
for $p=4$ in function of the temperature.


\section{The Short-Range Model}
\label{srsec}

Here we introduce and study a short-range version of the $p$-spin glass. 
The model that we introduce is 
defined on an hyper-cubic $d$-dimensional lattice of side $L$. 
On each side of the lattice there are $M$ spins. 
Every spin interacts via quenched random couplings with $p-1$ spins
chosen among spins on the same site and on nearest-neighbour
sites. It is natural to expect (and it will also derive from the equations)
that, for large $M$, one recovers the mean field solution since each spin 
interacts with a large number of other spins.
The Hamiltonian of the model is

\be
H(\{J\}) = \sum_{<l_1,\cdots,l_p>}^{L^{d}}
 \sum_{i_1,\cdots,i_p}^M \Jp s_{i_1}(l_1)\cdots s_{i_p}(l_p).
\ee

By $\sum_{<l_1,\cdots,l_p>}^{L^{d}}$ we sum over all the 
sites of the lattice taking, for each couple of adjacent sites $i$ 
and $j$, $p-k$ of the $l_1,\cdots,l_p$ indices equal to $i$ and 
$k$ indices equal to $j$. In other words, for each nearest neighbour
 sites $i$ and $j$, every interaction involves 
$p-k$ spins of site $i$ and $k$ spins of site $j$ with $k$ running 
from zero to $p$.
We consider discrete $(\pm 1)$ spin variables and we call $s_{i_r}(l_r)$ 
the $i_r^{th}$ spin of site $l_r$ with $i_{r}$ running 
from $1$ to $M$. 
The $\Jp$ are quenched random variables extracted from the
distribution

\be
P(\Jp) = \sqrt{\frac{a M^{p-1}}{\pi}}
 \exp{ \[[-a M^{p-1}(\Jp)^2 \]]},
\ee
where the normalization ensures an extensive free energy and 
the constant $a$ will be fixed later on by the condition that,
 in the large-$M$ limit, the free energy density of this model coincides
 with the one of the long range model.

The model just defined is a short range model, but 
the existence of the parameter $M$ ensures the mean field solution to 
be correct in the limit of $M \to \infty$. 
In this way one has a systematic way to reduce arbitrarily the
corrections to mean field which are due to the finiteness of the
coordination number and one can control the crossover effects from
 finite-dimensional to mean field behaviour. A similar generalization
 of the Random Energy Model has been studied in the one dimensional case
in \cite{cafrpa}. 

The mean field solution of the model is identical to the solution
of the long range $p$-spin model. 
Nevertheless, even in the large-$M$ limit, this model keeps
 an important ingredient of difference with respect to the 
 long-range $p$-spin model because there is a geometry in space.

The expression for the partition function of $n$ replicas of the system is

\be
\ov{Z^{n}} = e^{nML^d \beta^2/4} \prod_{a<b}^{n}\prod_{l}^{L}
 \int dQ_{ab}(l) d \lambda_{ab}(l)
 \exp\[[-M G[Q,\lambda] \]],
\label{zn1}
\ee
with 

\bq
G[Q ,\lambda] &=& - \frac{\beta^2}{2 a} \sum_{a<b}^{n} 
\sum_{<l_1,\cdots,l_p>}^{L^{d}}
\((Q_{ab}^{l_{1}}\ldots Q_{ab}^{l_{p}} \))^p + \sum_{a<b}^{n}
\sum_{l}^{L^{d}}
\lambda_{ab}^{l} Q_{ab}^{l} - \nonumber \\
&&  \ln \mbox{Tr} \exp
\[[ \sum_{a<b}^{n} \sum_{l}^{L^{d}} 
\lambda_{ab}^{l} s^{a}(l) s^{b}(l) \]],
\label{gpspinsr}
\eq

We can write the non local part of the $G[Q ,\lambda]$ in a more 
convenient form exploiting the following identity

\be
\sum_{<l_1,\cdots,l_p>}^{L^{d}}
\((Q^{l_{1}}\ldots Q^{l_{p}} \))^p = \sum_{l,m}^{L^{d}}
K(l,m) \((Q^{l} + Q^{m} \))^p - (2^{p} +2d -1) 
\sum_{l}^{L^{d}} (Q^{l})^{p},
\ee
where $$K(l,m) = \dd_{l,m} + \sum_{\vec{1}} \dd_{l+\vec{1},m},$$
and $ \sum_{\vec{1}}$ is the sum over all possible directions of the 
unitary lattice spacing vector $\vec{1}$.
So the function $G[Q ,\lambda]$ takes the form

\bq
G[Q ,\lambda] &=& - \frac{\beta^2}{2 a} \sum_{a<b}^{n} 
\sum_{l,m}^{L^{d}}
K(l,m) \((Q_{ab}^{l} + Q_{ab}^{m} \))^p - (2^{p} +2d -1) 
\sum_{l}^{L^{d}} (Q_{ab}^{l})^{p} \nonumber \\
&& + \sum_{a<b}^{n}
\sum_{l}^{L^{d}}
\lambda_{ab}^{l} Q_{ab}^{l} - \ln \mbox{Tr} \exp
\[[ \sum_{a<b}^{n} \sum_{l}^{L^{d}} 
\lambda_{ab}^{l} s^{a}(l) s^{b}(l) \]].
\label{gpspinsr1}
\eq

The mean field solution must be constant in 
space because the first term in (\ref{gpspinsr}) can be re-written as
 a gradient plus a local term.
We want to normalize the $G[Q ,\lambda]$ in such a way to recover 
(\ref{GGG}) in the mean field approximation. 
To do so one has to set 
$$a= 2^{p}d  -2d +1.$$

The saddle point equations yield

\bq
\lambda_{ab}^{l} = \frac{\beta^{2} p }{2}(Q_{ab}^{l})^{p-1} \nonumber \\
Q_{ab}^{l} = \left< S^{a}(l) S^{b}(l) \right>_{\lambda}.
\label{sppspinsr}
\eq

These equations are the same equations that we found in the long range 
case and are exact in the limit of $M \to \infty$.
Therefore, in the large-$M$ limit, the equations (\ref{sppspinsr}) 
reduce to equations (\ref{sprspspin})
or (\ref{sp1rsbpspin}) depending on the temperature. 

\section{Gaussian fluctuations}
\label{gausec}

For finite $M$ the saddle point equations are not exact. 
In this section we estimate the $O(1/M)$ corrections to the free energy due to
the small fluctuations around the saddle point of both the high and low 
temperature phase.
For finite $M$ we can write

\be 
Q_{ab} = Q_{ab}^{sp} + \dd Q_{ab}^l \hspace{1cm} 
\ll_{ab} = \ll_{ab}^{sp} + \dd \ll_{ab}^l ,
\ee
where $\dd Q_{ab}^l$ and $\dd \ll_{ab}^l$ are $O(1/M)$ quantities. 
$ Q_{ab}^{sp}$ and $\ll_{ab}^{sp}$ are the solutions
 of the equation (\ref{sppspinsr}) and are site independent.

\n
It is convenient to work in Fourier space so we write 

\be
\dd Q_{ab}^l =  \int_{-\pi}^{\pi} \frac{d^d \vec{k}}{(2 \pi)^d}
\dd Q_{ab}(k) e^{i \vec{k} \cdot \vec{l}}.
\ee

One can expand $G[Q ,\lambda]$ up to quadratic order in 
$\dd Q_{ab}^l$ and $\dd \ll_{ab}^l$. Observing that the couple
 $(\al \bb)$ takes $n(n-1)/2$ different values, we can write 

\be
\ov{Z^{n}} = e^{-M \beta F_{sp}} 
\prod_{\al=1}^{n(n-1)}
 \int \dd \OO_{\al} \exp \[[-\frac{M}{2} \sum_{\al<\bb}^{n(n-1)} 
 \int_{-\pi}^{\pi} \frac{d^d \vec{k}}{(2 \pi)^d}
\OO_{\al}(\vec{k})\hat{M}_{\al \bb}\OO_{\bb}(-\vec{k})\]],
\ee
where $MF_{sp}$ is the free energy of the system in the mean 
field approximation. We have used the following labeling:

\bq
\OO_{\al}(\vec{k}) &=& \dd \ll_{ab}(\vec{k}) \hspace{1cm} \mbox{for} 
\hspace{.5cm} 1 \le \al \le \frac{n(n-1)}{2} \nonumber \\
\OO_{\al}(\vec{k}) &=& \dd Q_{ab}(\vec{k}) \hspace{1cm} \mbox{for} 
\hspace{.5cm}  \frac{n(n-1)}{2} < \al \le n(n-1). \nonumber \\
\eq
 $\hat{M}_{\al \bb}$ is the $n(n-1)\times n(n-1)$ matrix of the fluctuations

\begin{equation}
\hat{M}=\left(
\begin{array}{c c }
M_{abcd}^{\ll,\ll}  & M_{abcd}^{q,\ll}  \\
M_{abcd}^{q,\ll} & M_{abcd}^{q,q} \\
\end{array}
\right),
\end{equation}
with
\begin{eqnarray}
 M_{abcd}^{q,q} &=&- \frac{ \beta^2 p (p-1)}{2} 
Q_{ab}^{sp (p-2)} \delta_{ab,cd} 
\frac{1-2d +2^{p}\cos^{2}\((\frac{{\bf k}}{2}\))}{1 - 2d +2^{p} d}
\nonumber \\
 M_{abcd}^{q,\ll} &=&
\delta_{ab,cd} ,  \nonumber   \\ 
 M_{abcd}^{\ll,\ll} &=& -  \[[ 
\left< S^{a}S^{b}S^{c}S^{d} \right>_{\lambda}
 -\left< S^{a} S^{b} \right>_{\lambda}
\left<S^{c} S^{d}\right>_{\lambda} \]] \equiv -C_{abcd}.
\label{fluct}
\end{eqnarray}
\noindent
 
We indicated with $C_{abcd}$ the connected four-spin correlation function.
We also used the notation $$\cos^{2}\((\frac{{\bf k}}{2}\)) \equiv 
\cos^2\((\frac{k_{1}}{2}\))+\ldots+\cos^2\((\frac{k_{d}}{2}\)).$$

Performing the Gaussian integral one obtains

\be
F(T) = F_{sp}(T) + \frac{1}{M} \Delta F(T),
\ee 
with

\be
\Delta F=  \frac{1}{2 \beta}
\lim_{n \to 0}
 \sum_{\lambda} 
m_{\lambda} \int_{-\pi}^{\pi}\frac{d^d{\bf k}}{(2\pi)^d}
\ln \[[ \lambda({\bf k })\]],
\ee
where $m_{\lambda}$ is the multiplicity of the 
$\lambda(\vec{k})$ eigenvalue of $\hat{M}_{\al \bb}$.

If the temperature is above the critical temperature 
the mean field solution is RS with $q_{0}=0$ (we shall always 
consider the case of zero magnetic field).
The matrix of the fluctuations has got the whole block $M_{abcd}^{q,q}$
equal to zero and it is straightforward to show that above $T_c$ 
the temperature-dependent 
part of the $O(1/M)$ corrections is zero and

\be
\Delta F =0.
\label{dgat}
\ee

In the low temperature phase the structure of $\hat{M}_{\al \bb}$
changes because the sub-matrix $M_{abcd}^{q,q}$ is not identical to 
zero but takes the following form
\bq
M_{abcd}^{q,q}&=& 0  \nonumber \mbox{
       if $a$ and $b$ do not belong to the same block } \\
M_{abcd}^{q,q} &=& f(q,{\bf k})
\mbox{      if $a$ and $b$ belong to the same block }, \nonumber
\eq
where

\be
f(q,{\bf k}) = - \frac{ \beta^2 p (p-1)}{2} q^{p-2}
\frac{1-2d +2^{p}\cos^{2}\((\frac{{\bf k}}{2}\))}{1 - 2d +2^{p} d}.
\label{reffq}
\ee

In this case one needs to calculate all
 the $n(n-1)$ eigenvalues of $M_{\al,\bb}$. 
The calculation is similar to the one performed in \cite{bpr}
 in the case of
 the Little model with the difference that in that case the model was
 Gaussian and there was no need to introduce the $\lambda_{\al \bb}$ field.
 So here the eigenvalues are $n(n-1)$.
We shall indicate with $\lambda_{\Lambda}$ and  $\lambda_{Q}$ the eigenvalues 
of the two sub-blocks $M_{abcd}^{\lambda,\lambda}$ and $M_{abcd}^{q,q}$.     
It is easy to calculate the determinant of the total matrix $\hat{M}$ observing
that, because of the diagonal structure of $M_{abcd}^{q,q}$,
 correspondingly to every $\lambda_{\Lambda}$ and $\lambda_{Q}$
there are two eigenvalues of the total matrix 
\be
\lambda_{\pm} = \frac{\lambda_{\Lambda} + \lambda_{Q} \pm
\sqrt{(\lambda_{\Lambda} - \lambda_{Q})^2 +4}}{2},
\ee
so that $\lambda_{+}\lambda_{-} = \lambda_{\Lambda} \lambda_{Q}-1 $.

The structure of the spectrum of the eigenvalues  $\lambda_{\Lambda}$
 and  $\lambda_{Q}$ in the 1RSB phase is 
briefly reported in the appendix.
The final expression for the $O(\frac{1}{M})$ corrections below 
$T_{c}$ becomes

\bq
&&\Delta F = 
\frac{1}{2\beta} \int_{-\pi}^{\pi}\frac{d^{d}{\bf k
}}{(2\pi)^{d}} 
\left\{ \frac{(m-3)}{2} \ln \[[ 1 + f(q,{\bf k}) \((C_{[abcd]} +
(1-q)^{2}\))
 \]] \right. \nonumber \\
 &&\frac{1}{m} \ln \[[ 1 - f(q,{\bf k})  \((
\frac{A + D + |A-D|}{2} \)) \]]  \nonumber \\
&& \left. \frac{m-1}{m} \ln \[[ 1 - f(q,{\bf k}) \((
\frac{A' + D' + |A'-D'|}{2} \))  \]] \right\},
\label{corr1}
\eq

\n
where we have defined 
$$C_{[abcd]}=\left< S^{a}S^{b}S^{c}S^{d} \right>_{\lambda}
 -\left< S^{a} S^{b} \right>_{\lambda}
\left<S^{c} S^{d}\right>_{\lambda},$$ with the four replicas $a,b,c,d$ 
belonging to the same diagonal block of the 1RSB matrix.
The constants $A,D,A',D'$ in equation (\ref{corr1}) 
are also defined in the appendix.

\begin{figure}[htpb]
\centerline{\epsfig{figure=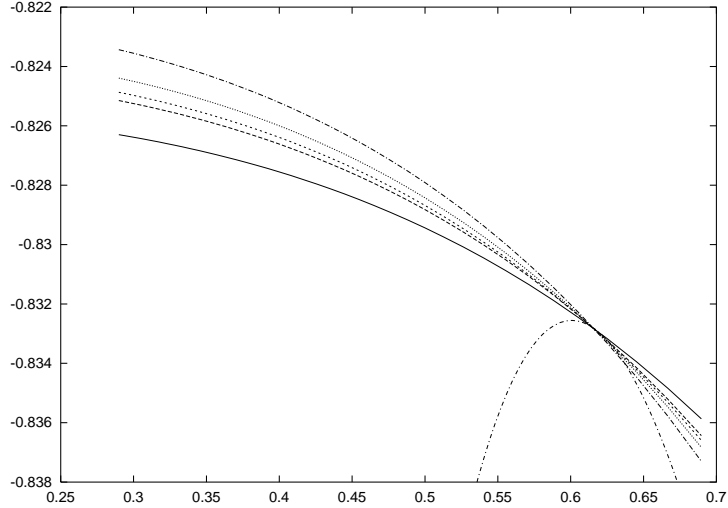,angle=270
,width=10cm}}
\vspace{0cm}
\caption[]{Free energy density of the three-dimensional 
$p$-spin model vs temperature for $p=4$.
 and $M=4,6,8,10,\infty$}
\label{fM}
\end{figure}

As a check one can easily verify that, 
for $m=1$, the result (\ref{dgat}) is, as it should be, recovered since 
the structure of the equations becomes identical to the RS case after 
the identification $q_{1}=q_{0}$.

To perform the above integrals in momentum space
 we make use of the identity

\be
\int_{-\pi}^{\pi}\frac{d^{d}{\bf k}}{(2\pi)^{d}} 
\ln\[[1+A{\cos^{2}(\frac{{\bf k}}{2})}\]] 
= - \int_{0}^{\infty}
\frac{dt}{t} \exp[-t] \[[I_{0}^d \((\frac{A  t}{2}\))E^{-\((\frac{A d t}{2}
\))} -1\]],
\ee

\n
where $I_0$ is the modified Bessel function of order $0$. 

One can solve numerically the saddle point equations and then 
plug the value of
$q_1(T),m(T)$ into the expression of $G$ and $\Delta G$. 
The result, for $d=3$ and $M=4,6,8,10,\infty$, is shown
 in figure (\ref{fM}). In the plot the higher are the curves the smaller
is the value of $M$. 
We  can see that the perturbative corrections slightly shift the curves
but do not change the nature of the transition. The critical 
temperature, defined as the temperature where the 
curves of the high and low temperature free energy coincide, is the same 
at which $m_{sp}=1$. Furthermore, the two curves are tangent at the critical 
temperature denoting a zero latent heat as in the MF case.
Finally, we also remark
that the transition temperature changes very slightly with $M$. A rough
estimate the $O(1/M)$ shift in the critical temperature gives
$$T_c(M) \simeq T_c -\frac{1}{M} (0.0168).$$

\section{Replica Propagators}
\label{prosec}
\subsection{At equilibrium}

The study of the Gaussian propagators provides knowledge on the existence
 of zero-mass modes and therefore on 
the nature of the transition.
In this section we consider the corrections to the 
mean field solution of the theory defined by the (\ref{gpspinsr1}). 
In a perturbative approach we define and 
calculate the propagators of the $p$-spin model in the 
RS and 1RSB phase.   

The propagators $G_{abcd}$ are defined in the following way
\bq
G_{abcd}^{Q}&&= \la \dd Q_{ab} \dd Q_{cd} \ra \nonumber \\
G_{abcd}^{ \La Q}&&\equiv G_{cdab}^{ Q \La }
= \la \dd \La_{ab} \dd Q_{cd} \ra \nonumber \\
G_{abcd}^{\La}&&= \la \dd \La_{ab} \dd \La_{cd} \ra. \nonumber
\eq

The equations for the propagators are easily found by standard arguments

\be
\la \phi(x) \frac{\partial{S[\phi]}}{\partial{\phi(y)}} \ra
= \dd(x-y),
\label{eqprogen}
\ee
and in the case of our functional integral (\ref{zn1}) they become

\bq
\la \dd Q_{ab} \frac{\partial{G[Q,\La]}}{\partial{Q_{cd}}} \ra
&=& \dd_{ab,cd} \hspace{1cm}
 \la \dd \La_{ab} \frac{\partial{G[Q,\La]}}{\partial{Q_{cd}}} \ra
= 0 \nonumber \\
\la \dd Q_{ab} \frac{\partial{G[Q,\La]}}{\partial{\La_{cd}}} \ra
&=& 0 \hspace{1cm}
 \la \dd \La_{ab} \frac{\partial{G[Q,\La]}}{\partial{\La_{cd}}} \ra
= \dd_{ab,cd}.
\label{eqpro}
\eq

It is straightforward to obtain the following linear
 equations for the propagators 
\bq
&&2 f(Q_{cd},{\bf k})
 G_{abcd}^{Q} + 2 G_{abcd}^{\La Q}  = \dd_{ab,cd} \nonumber \\
&&2 f(Q_{cd},{\bf k})
 G_{abcd}^{\La Q} + 2 G_{cdab}^{\La}  = 0 \nonumber \\
&&2 G_{abcd}^{Q} - 2 \sum_{r<s} C_{r s c d}
 G_{ r s a b}^{\La Q}   = 0 \nonumber \\
&&2 G_{abcd}^{\La Q} - 2 \sum_{r<s} C_{r s cd}
 G_{ r s a b}^{\La }   =  \dd_{ab,cd},
\label{props}
\eq
where $ f(Q_{cd},{\bf k})$ is defined by (\ref{reffq}).

We shall always use the fact that, in absence of magnetic field, $q_{0}=0$.
There are different kind of propagators depending on the relation between the 
replica indices they refer to, so for different choices of indices 
$a b c d$ in equations (\ref{props}) we will obtain different 
equations.
In the RS phase one obtains 
\bq
 &&G_{a b c d}^{\La} =0 \\
&& G_{a b c d}^{\La Q} =
 \frac{1}{2} \dd_{a b,c d} \\
&& G_{a b c d}^{Q} =
 \frac{1}{2} C_{a b c d},
\label{propsrs}
\eq
where $C_{a b c d}$ is the four-spin connected correlation function
defined in (\ref{fluct}). So in the replica-symmetric phase all propagators 
$ G_{a b c d}^{Q}$ are zero except for the diagonal one which is trivial 
\be
 G_{a b a b}^{Q} =  G_{a b b a}^{Q} =\frac{1}{2}.
\ee

In the 1RSB phase, we shall use the convention to indicate with $[a b]$ two 
replicas belonging to the same block and with $[a][b]$ two replicas 
belonging to different blocks of the
1RSB saddle point matrix $Q^{sp}_{ab}$. If $a$ and $b$ (or $c$ and $d$) 
do not belong to the same block , the form of the propagators is still 
given by equation (\ref{propsrs}) where the function $C_{a b c d}$
changes in the low temperature phase so one has

\bq
&&G_{01}^Q \equiv G_{[a] [b ][a][b ]}^{Q} = \frac{1}{2}. \nonumber \\
&&G_{02}^Q \equiv G_{[a] [b ][c][b ]}^{Q} = \frac{1}{2} q_{1}, \ \ \ [ac]. 
\nonumber \\
&&G_{03}^Q \equiv G_{[a] [b ][c][d ]}^{Q} = \frac{1}{2}q_{1}^{2}, \ \ \ 
[ac]\ \ \ [bd].  
\eq
where $a$ and $b$ belong respectively to the same blocks as $c$ and $d$. 

For the replica indices in the same block, $ [a b]$, 
we have $Q_{ab}=q_1$ and from equations (\ref{props}) we obtain the 
following equations for $G^{\La}$ and $G^{Q}$

\bq
G_{a b c d}^{ Q} &=&  f(q_1,{\bf k})^{-1} \dd_{a b , c d}
 + f(q_1,{\bf k})^{-2}
 G_{ a b c d}^{ \La} \nonumber \\
G_{ a b c d}^{ \La} &=& - f(q_1,{\bf k}) \sum_{r<s} C_{r s c d}
 G_{ r s a b}^{\La } -\frac{f(q_1,{\bf k})}{2}\dd_{a b, c d} . 
\label{eqspropa}
\eq

\begin{figure}[htpb]
\centerline{\epsfig{figure=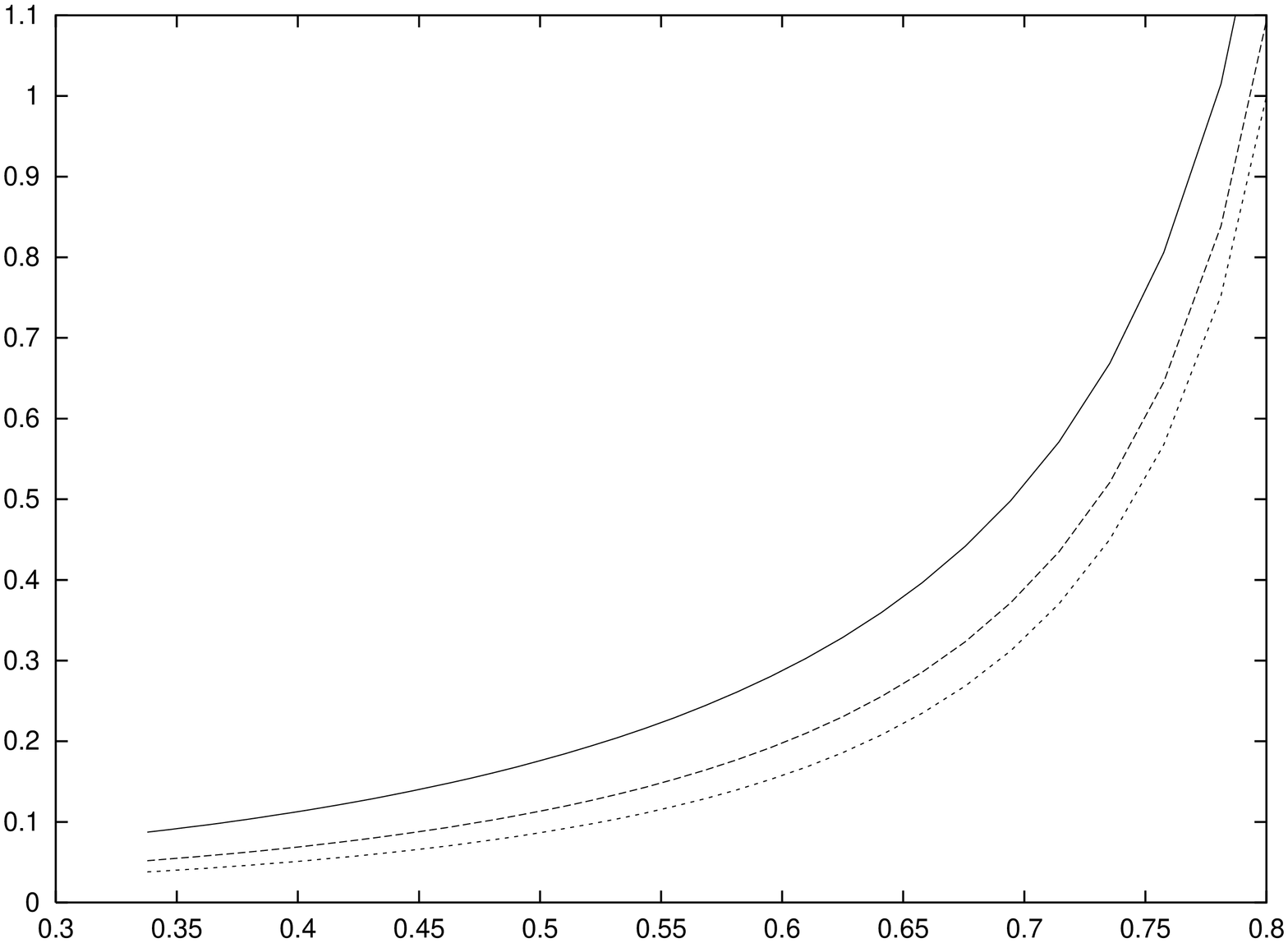,angle=270
,width=10cm}}
\vspace{0cm}
\caption[]{$G_{1}^{\La}$, $G_{2}^{\La}$, $G_{3}^{\La}$ for $p=4$ and $d=3$}
\label{prop123ps}
\end{figure}

The second of the equations (\ref{eqspropa}) can be written
for the various choices  
of the indices $a b c d$ obtaining a set of four coupled equations for 
$G_{1},G_{2},G_{3},G_{4}$ defined as follows:

\bq
G_{1} &=&  G_{[a b a b]} \hspace{1cm} G_{2} = 
 G_{[a b a r]} \nonumber \\
G_{3} &=&  G_{[a b r s]} \hspace{1cm} G_{4} = 
 G_{[a b][r s]} 
\eq

Solving the equations and plugging in the MF values for  
$q_1$ and $m$ we obtain the propagators in function of the temperature. 
From the first of the equations (\ref{eqspropa}) it is possible to define 
the propagators $G^Q_{abcd}$. 
One has that, in absence of magnetic field $G_{4}^{\La}(k=0)=0$.
In figure (\ref{prop123ps}) we plot the 
$G_{1}^{\La},G_{2}^{\La},G_{3}^{\La}$ for $k=0$ versus the temperature.
In conclusion, we observe no divergence of the RS and 1RSB propagators
at the critical temperature.

This means that no zero-mass 
modes are present around the stable saddle point solutions and usually 
it indicates that no continuous transition is taking place. 

\subsection{On the dynamical line}
We recall that in the long range $p$-spin model, there 
is a dynamical critical temperature $T_d > T_c$ below which the system does
not reach equilibrium in finite times. 
In the mean field approach, if we quench the system to low temperature
coming from an high temperature state, the system goes to a dynamical 
metastable state, having an energy greater than the equilibrium one.
In \cite{kucu}, it has been shown  that if the system starts
from a random initial configuration, it evolves following the  
flat directions connecting an ensemble of metastable states which are the 
threshold solutions to the TAP equations \cite{culde}.

This cannot happen in short range models and it is and artifact
of mean field theory. However it can be interpreted as the signal 
of starting a quite slow approach to equilibrium. In any case the 
disappearance of this metastable energy is a non perturbative result, so that 
it make sense to study the properties of the correlation functions in this 
state.

In the replica approach it can be shown that in the dynamical metastable
state the value of $m$ is not fixed my extremizing the free energy, but 
by imposing the condition that the replicon 
eigenvalue is equal to zero.
We are going to compute the propagators in this region as functions of the
temperature. The formulae are similar to the previous ones, only the value
of $m$ will be different.

This amounts to set
\be
f(q,{\bf k=0}) = \frac{1}{\lambda_{\Lambda}^{R}}
\ee
where $\lambda_{\Lambda}^{R}$ is defined in the appendix.

We thus obtain some new equations for the propagators concerning replicas
belonging to the same block 

\begin{eqnarray}
G_1=\frac{(6 -5m+m^2)(q-r_0)(4q-m q^2-4r_0+mr_0)}{A
(f(q,{\bf k})-l_0)
(f(q,{\bf k})-l_0-(m-1)l_1)(f(q,{\bf k})-l_2)}\\
G_2=\frac{(3-m)(q-r_0)(4q-m q^2-4r_0+mr_0)}{A (f(q{\bf k},)-l_0)
(f(q,{\bf k})-l_0-(m-1)l_1)(f(q{\bf k},)-l_2)}\\
G_3=\frac{2 (q-r_0)(4q-m q^2-4r_0+mr_0)}{A(f(q,{\bf k})-l_0)
(f(q,{\bf k})-l_0-(m-1)l_1)(f(q{\bf k},)-l_2)}
\label{margprop}
\end{eqnarray}
where
\begin{eqnarray}
A &=&                                                  
(-1 + 2 q - r_0)^2 (-2 + 16 q - 6 m q - 32 q^2  + 23 m^2 q  - 
   3 m^2  q^2  + 4 m q^3  - 5 m^2  q^3  + \nonumber \\
&&  m^3  q^3  - 12 r_0 + 7 m r0 -
  m^2  r_0 + 48 q r_0 - 46 m q r_0 + 13 m^2  q r_0 +\nonumber \\
&&- m^3  q r0 - 
3 m q^2  r_0 + 4 m^2  q^2  r_0 - m^3  q^2  r_0 - 18 r_0^2  + 21 m r_0^2  - 
8 m^2  r_0^2  + m^3  r_0^2 ) \nonumber \\
l_0&=&\frac{-1}{1-2q+r_0} \nonumber \\
l_1&=&\frac{1}{-1+4 q-m q-3 r_0+ m r_0} \nonumber \\
l_2&=&\frac{-2}{2-8 q+4 m q +m q^2- m^2 q^2+6 r_0 -5 m r_0+m^2 r_0}.
\end{eqnarray}

In the previous formulae, $r_0=C_{[abcd]}$. 
In the general case the propagators diverge, for small $k$, as $k^{-2}$. 
The condition $m=1$ now gives the dynamical critical temperature, 
and one can see that all the propagators in (\ref{margprop})
 coincide and that the divergence is of order $k^{-4}$.

The divergence at small momenta of the propagator is not  unexpected 
\cite{FPcom}. Indeed the
vanishing of the replicon propagator implies a divergence of the susceptibility
and consequently a singularity at $k=0$.
The form of the singularity could not be predicted using general arguments.
The change of the exponent at $T_d$ is particularly striking; it is related
to the degeneracy of the replicon and longitudinal eigenvalue at $m=1$.

Although the result was derived in the context of spin models, we believe
that this structure of exponents $k^{-4}$ at $T_d$ and $k^{-2}$ at $T<T_d$
is quite general and it would be valid in many others models. 
It is also clear that these are mean field results: also if we remain in
a perturbative framework, the exponents are likely to be changed for sufficient
low dimension. The value of the upper critical dimension (6?, 8?), above
which the mean fields exponents do not get perturbative corrections, can
be extracted by analyzing the contribution of higher loops, but this
task goes beyond the aim of this paper.

\section{Discussion}
\label{numsec}

The model that we have described has been studied numerically in
$d=3$ with $p=4$ and $M=3,4$ for $T>T_c$ \cite{frpa} and below 
$T_c$ \cite{cacopa}.
The results of the numerical simulations seem to indicate,
the existence of a transition at a critical temperature $T_{c}$
from a high temperature phase to a broken replica symmetry phase.
However, this transition appears to be of second order with divergent 
spin-glass susceptibility. 
A possible interpretation of this apparent contradictory phenomenology
is the following \cite{KTW}.
For each realization of the disorder there are some regions of space in
 which the effect of frustrations are weaker than in the rest of the system. 
Within these regions the system is likely to freeze at a temperature $T_r$
which is higher than the temperature $T_c$ at which the whole system freezes.
So for $T>T_c$ (not too high temperatures though!) there are regions in
 space which the system is locally frozen,
 the typical size of the region being a function of the temperature 
$R^{d_r}(T)$ where $d_r$ is the dimensionality of the region.
Within these regions the system is very strongly correlated and the
 correlation length is of order $R(T)$.
The total SG susceptibility is the integral over space of the local 
SG susceptibilities and the contribution of the regions
 in which the system is strongly
 correlated grows with $R(T)$ and diverges when $R(T)$
 becomes of the size of the whole system.
 So, in this interpretation, the transition remains
 discontinuous within the regions of space where it occurs, {\i.e.} the
local overlap changes discontinuously from $q_0$ to $q_1$, and the continuous
varying quantity is the typical size of the regions where the system is 
strongly correlated.
This work wants to be
 a step forward towards the comprehension of spin glass 
models and of structural glasses in finite dimensions parallel to other
 attempts \cite{cacopa,frpa,papiri}. There are still many unclear things on
 the subject which is worth, according to us, for further studies.

\section{Acknowledgments}
We are happy to thank A. Cavagna and F. Ritort for useful discussions.

\appendix

\section{}

In the 1RSB phase there are three invariant subspaces of eigenvectors 
corresponding to three classes of eigenvalues. In this appendix we 
shall list the eigenvalues $\lambda_\lambda$ and $\lambda_Q$ 
and their multiplicity $\mu$ for each different subspace. 

\subsection{Longitudinal eigenvalues}

For each diagonal sub-matrix there are two couples of 
longitudinal eigenvalues with multiplicity $ \mu_L=1$

$$\lambda_{\Lambda}^{LA} = \frac{A + D \pm |A-D|}{2}  
\hspace{1cm} \lambda_{Q}^{LA} = f(q_1,{\bf{k}}) $$
 
where

\bq
A && = 1-q_1^2 + 2(m-2)(q_1 - q_1^2) - \frac{(m-2)(m-3)}{2} C_{[abcd]} \\
D &&= -1 +2(m-1)q_1 + (m-1)^2 q_1^2 
\eq

\subsection{Anomalous eigenvalues}

For each diagonal sub-matrix there are four couples of 
anomalous eigenvalues.

Two of them have multiplicity $\mu_A=(n-m)/m$ and coincide with the
longitudinal ones.

The other two have multiplicity $\mu_A=n(m-1)/m$ and are 
$$\lambda_{\Lambda}^{LA} = \frac{A' + D' \pm |A'-D'|}{2}  
\hspace{1cm} \lambda_{Q}^{LA} = f(q_1,{\bf{k}}) $$
 
where

\bq
A' && = 1-q_1^2 + (m-4)(q_1 - q_1^2) - (m-3) C_{[abcd]} \\
D' &&= -1 +(m-2)q_1 - (m-1) q_1^2 
\eq

\subsection{Replicon eigenvalues}

There are four couples of different replicon eigenvalues

\bq
\lambda_{\Lambda}^{R} = P_1 -2Q_1 +R_0 &&\hspace{1cm}\lambda_{Q}^{R} = 
f(q_1,{\bf{k}}),
 \hspace{1truecm} \mu_{R} = n\frac{(m-3)}{2} \nn \\
\lambda_{\Lambda}^{R} = P_0 + 2(m-1)Q_0 +(m-1)^2 q_1^2
  &&\hspace{1cm}\lambda_{Q}^{R} = 0 ,\hspace{1truecm} \mu_{R} = n\frac{(n-3m)}{2m^2} \nn \\
\lambda_{\Lambda}^{R} =  P_{0} +(m-2) Q_{0} - (m-1)q_{1}^2
  &&\hspace{1cm}\lambda_{Q}^{R} = 0 ,
 \hspace{1truecm} \mu_{R} = n\frac{(n-2m)(m-1)}{m^2} \nn \\
\lambda_{\Lambda}^{R} =  P_{0} -2 Q_{0} +q_1^2
  &&\hspace{1cm}\lambda_{Q}^{R} = 0 ,
 \hspace{1truecm} \mu_{R} = n\frac{(n-m)(m-1)^2}{2m^2} 
\eq
where
\bq
P_0 &=& -1 \hspace{1cm} P_1  = q_1^2 -1 \nn \\
Q_0 &=&-q_1  \hspace{1cm} Q_1 = q_1^2 - q_1  \nn \\
R_0 &=& q_1^2 - C_{[abcd]} 
\eq

\end{document}